\documentclass[reprint,twocolumn,superscriptaddress,groupedaddress,showpacs,aps,prl]{revtex4}

\usepackage{graphicx}
\usepackage{dcolumn}
\usepackage{bm}

\begin{document}

\title{Effects of Hyperbolic Rotation in Minkowski Space\\ on the Modeling of Plasma Accelerators in a Lorentz Boosted Frame}

\author{J.-L. Vay}
\email{jlvay@lbl.gov}
\author{C. G. R. Geddes}
\affiliation{Lawrence Berkeley National Laboratory, Berkeley, CA, USA\\}

\author{E. Cormier-Michel}
\affiliation{Tech-X Corporation, Boulder, CO, USA\\}

\author{D. P. Grote}
\affiliation{Lawrence Livermore National Laboratory, Berkeley, CA, USA\\}

\date{\today}
\begin{abstract}
Laser driven plasma accelerators promise much shorter particle accelerators but their development requires detailed simulations that challenge or exceed current capabilities. We report the first direct simulations of stages up to 1~TeV from simulations using a Lorentz boosted calculation frame resulting in a million times speedup, thanks to a frame boost as high as $\gamma=1300$. Effects of the hyperbolic rotation in Minkowski space resulting from the frame boost on the laser propagation in the plasma is shown to be key in the mitigation of a numerical instability that was limiting previous attempts.

\end{abstract}


\maketitle


Laser driven plasma waves produce accelerating gradients orders of magnitude greater than standard accelerating structures (which are limited by electrical
breakdown) \cite{TajimaPRL79,EsareyRMP09}. 
High quality electron beams of energy up-to 1 GeV 
have been produced in just a few centimeters \cite{GeddesNature04,ManglesNature04,FaureNature04,LeemansNature06} 
with 10 GeV stages being planned as modules of a high energy collider \cite{SchroederAAC08}, and detailed simulations are required to realize the  promise of much shorter particle accelerators using this technique \cite{BruhwilerAAC08}.  Such  simulations challenge or exceed current capabilities, in particular for high energy stages at GeV energies and beyond.  

The linear theory predicts that for the intense lasers (a$\gtrsim$1) typically used for acceleration, the laser depletes its energy over approximately the same length $L_d=\lambda_p^3/2\lambda_0^2$ over which the particles dephase from the wake, where  $\lambda_p=\sqrt{\pi c^2m/e^2n_e}$ is the plasma wavelength, $\lambda_0$ is the laser wavelength, $c$ is the speed of light, and $m$, $e$ and $n_e$ are respectively the electron mass, charge and density in the plasma \cite{TajimaPRL79}. 
As a result of beam dephasing and laser depletion, the maximum bunch energy gain scales approximately as the square of the plasma wavelength and the inverse of the plasma density, 
which implies that higher energy stages operate with longer plasmas, rending computer simulations more challenging, as the ratio of longest to shortest spatial lengths of interest (plasma length/laser wavelength) rises. As a matter of fact, direct explicit multi-dimensional simulations of 10 GeV stages, which will operate in m-scale plasmas at order $10^{17}/cc$ densities, have been considered until recently beyond the current state of the art \cite{BruhwilerAAC08,GeddesPAC09}.

Recently, first principles Particle-In-Cell modeling of laser-plasma wakefield accelerators using a Lorentz boosted frame of reference \cite{VayPRL07} have been shown to being sped up by up-to three orders of magnitude in the calculations of stages in the 100 MeV-10 GeV energy range \cite{BruhwilerAAC08,VayPAC09,MartinsPAC09,VaySciDAC09,HuangSciDAC09,VayDPF09,MartinsCPC10,MartinsPoP10,MartinsNP10}. 
Maximum obtainable speedups calculated using linear theory predict that higher speedups are attainable, in the range of 4-6 orders of magnitude for stages in the energy range of 10 GeV-1 TeV respectively  \cite{VayAAC10,VayARXIV10}. Practical limitations have prevented reaching these speedups, including a violent high frequency numerical instability,  limiting the Lorentz boost $\gamma$ below $100$ \cite{BruhwilerAAC08,VaySciDAC09,MartinsCPC10,VayAAC10,VayARXIV10}. 

We report for the first time direct explicit simulations of stages in the range of 0.1 GeV-1 TeV, using a Lorentz boosted calculation frame with gamma as high as 1300,
verifying the performance and energy gain scaling \cite{CormierAAC08,GeddesPAC09} of plasma accelerator stage with deep laser depletion into the 1 TeV range and providing the tools for detailed designs for upcoming 10 GeV experiments such as BELLA \cite{LeemansAAC10}.   
As we have shown in \cite{VayPRL07}, the speedup provided by computing using a Lorentz boosted frame comes from the properties of space and time contraction and dilation of the Lorentz transformation. In this paper, the property of rotation of space-time of the Lorentz transformation is utilized to overcome the numerical instability that has arisen for boost values needed for reaching the maximal theoretical speedup.
In conjunction with the development of novel numerical techniques that are described elsewhere \cite{VayAAC10,VayARXIV10}, this allows the simulations to approach the theoretically calculated speedups of  4-6 orders of magnitude for 10 GeV-1 TeV stages, which in turn allows simulations of high energy plasma accelerators.

\begin{figure}[htb]
\centering
{\small
\begin{tabular}{@{}c@{}c@{}}
  Lab ($\gamma=1$) &
  Boost ($\gamma=5$) \\
  \includegraphics*[height=0.19\textwidth]{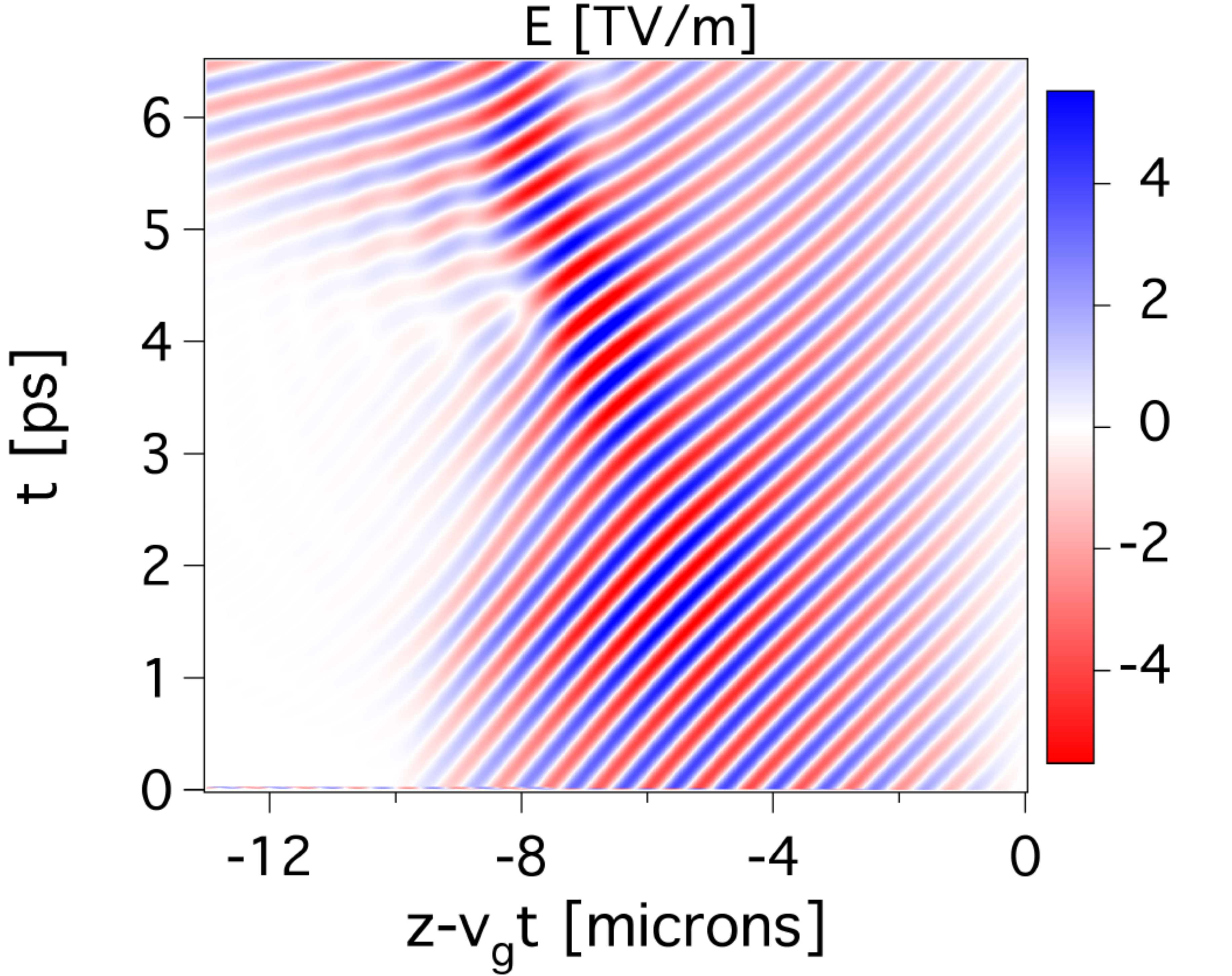} &
  \includegraphics*[height=0.19\textwidth]{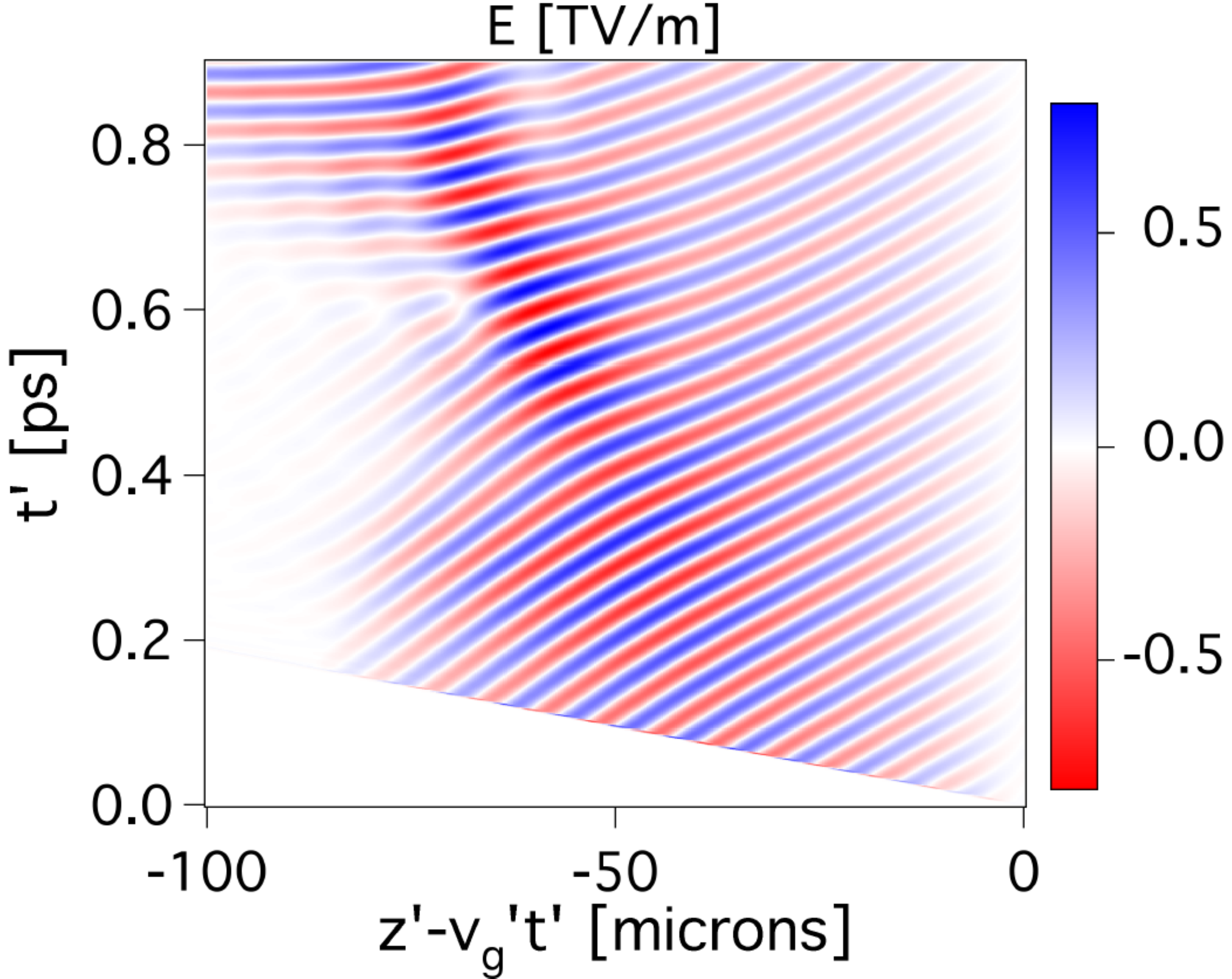} \\
  Boost ($\gamma=13$) &
  Boost ($\gamma=13$) in Lab \\
  \includegraphics*[height=0.19\textwidth]{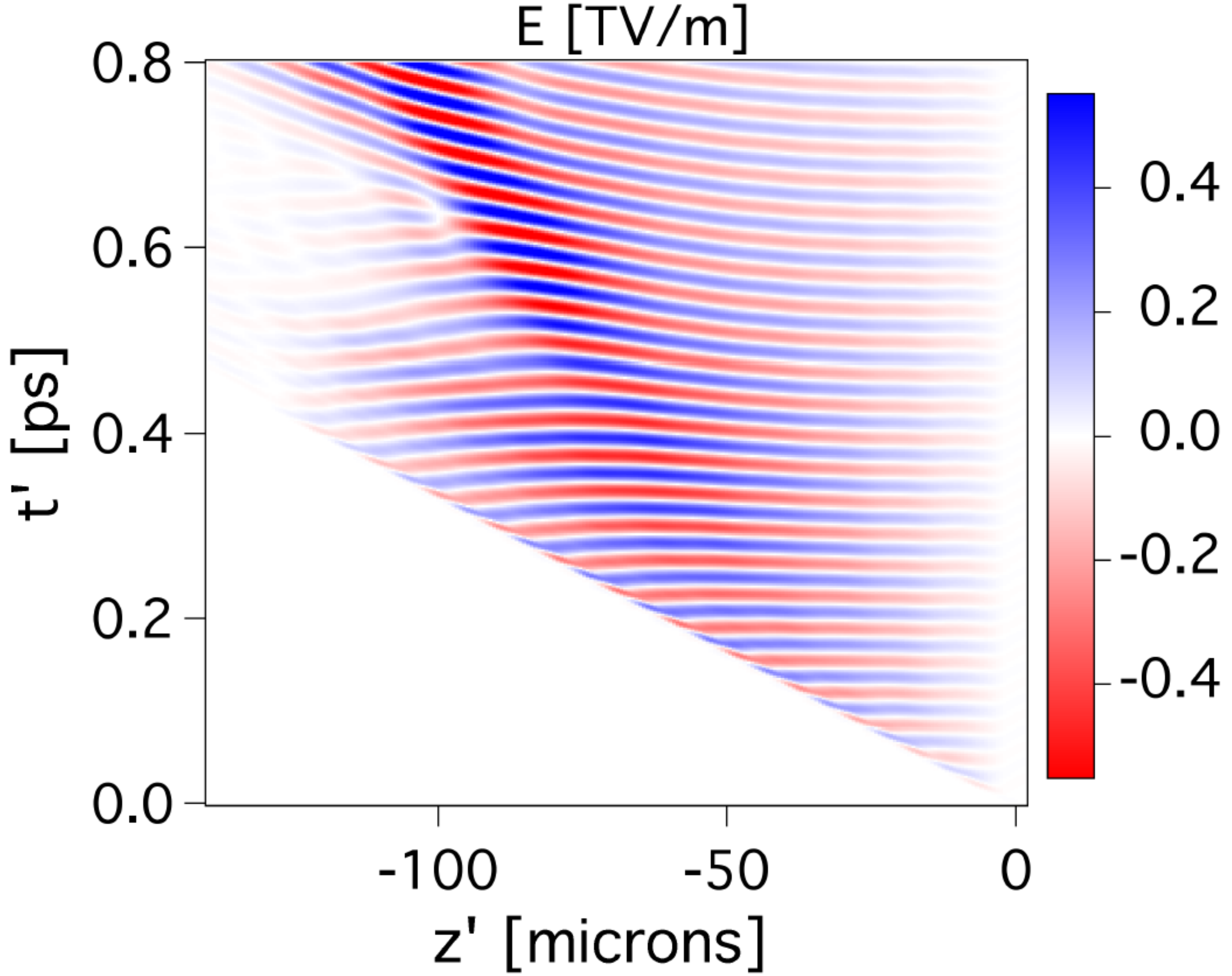} &
  \includegraphics*[height=0.19\textwidth]{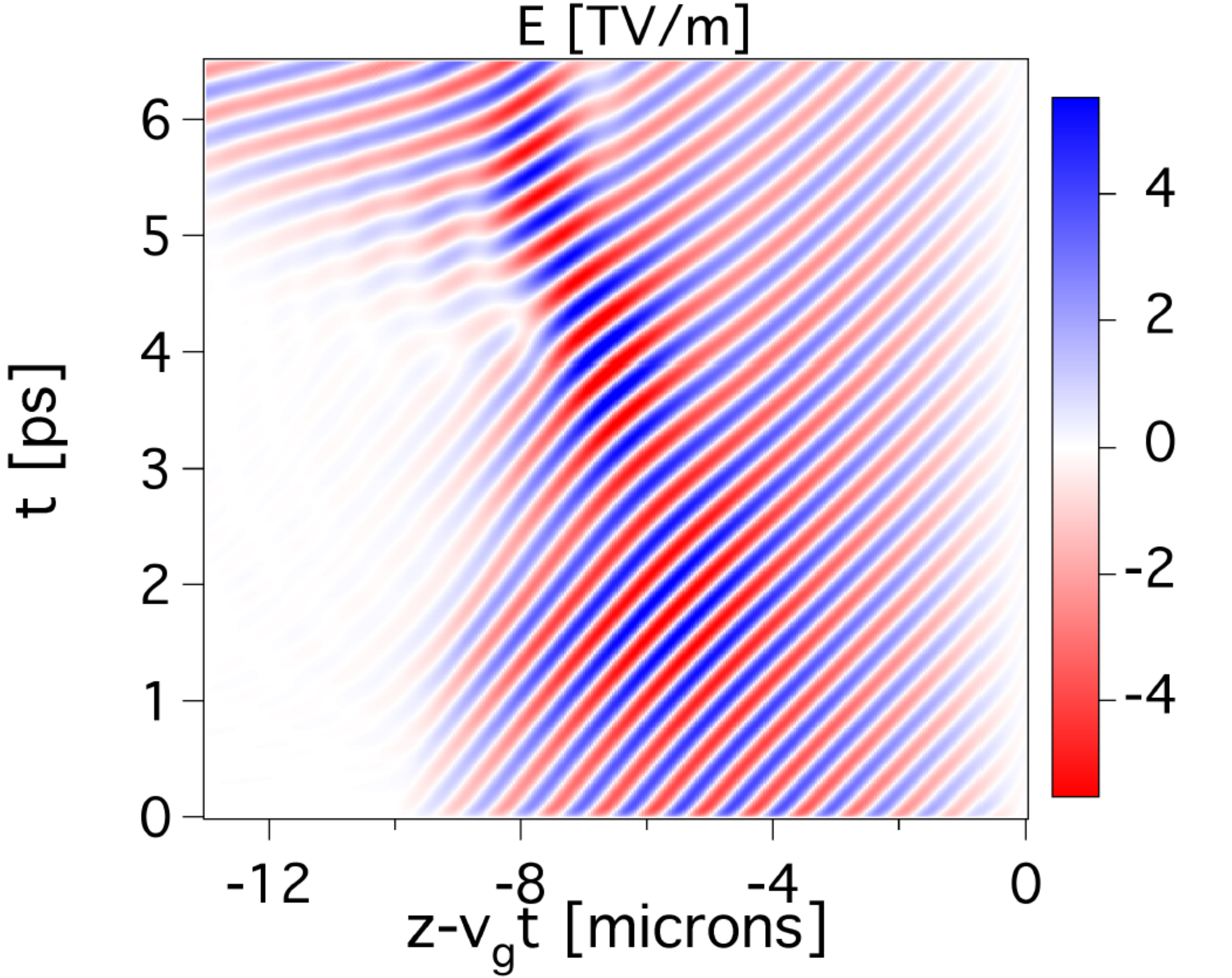} 
\end{tabular}
}
  \caption{(color online) Color plots of the laser field history on axis in a moving window propagating at the laser group velocity from simulations: (top-left) in the laboratory frame; (top-right) in a boosted frame at $\gamma=5$; (bottom-left) in a boosted frame at $\gamma=13$; (bottom-right) in a boosted frame at $\gamma=13$ plotted in laboratory coordinates after Lorentz transformation of the data.}
\label{Fig_hyprot}
\end{figure}

\textit{ Effect of the hyperbolic rotation in Minkowski space.---}
\begin{figure*}[htb]
    \centering
\begin{tabular}{m{1.8cm}m{7.cm}m{7.cm}}
  &
  \makebox[7.cm]{Lab ($\gamma=1$)} &
  \makebox[7.cm]{Boost ($\gamma=13$)} \\
   {\large Laser} &
    \includegraphics*[width=70mm]{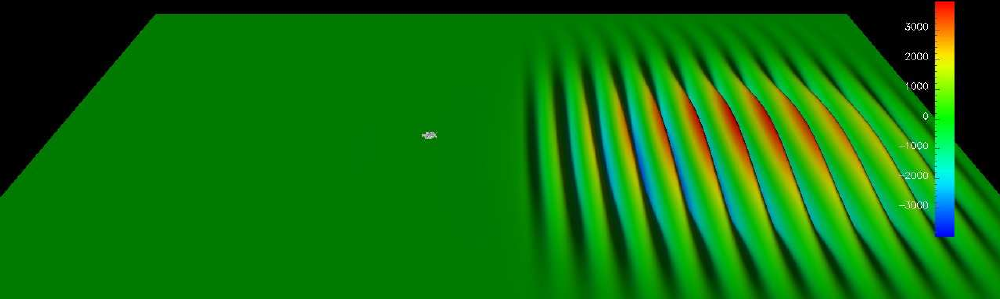} &
    \includegraphics*[width=70mm]{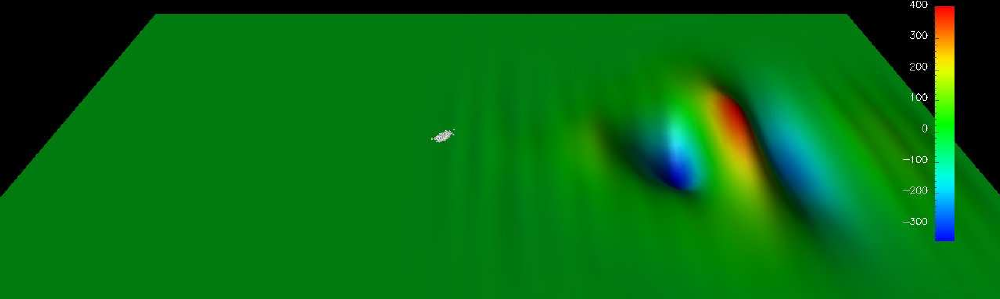} \\
   {\large Wake} &
    \includegraphics*[width=70mm]{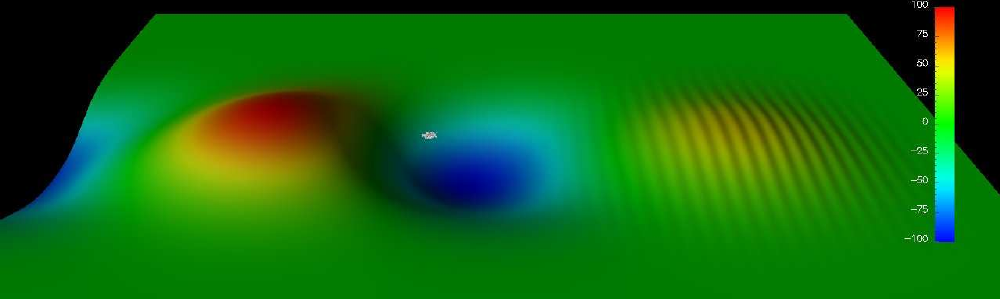} &
    \includegraphics*[width=70mm]{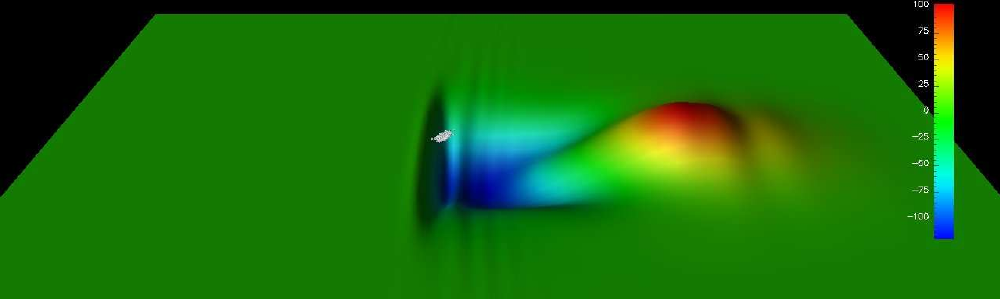}
\end{tabular}
    \caption{(color online) Colored surface rendering of the transverse (laser) and longitudinal (wake) electric fields from a 2-1/2D Warp simulation of a laser wakefield acceleration stage in the laboratory frame (left) and a boosted frame at $\gamma=13\approx\gamma_{wake}$ (right),  with the beam (white) in its early phase of acceleration. The laser and the beam are propagating from left to right.}
    \label{Fig_surf2de}
\end{figure*}
The effects of the Lorentz transformation on the laser and wake propagation through a 100 MeV laser plasma acceleration stage \cite{CormierAAC08,GeddesPAC09}
is illustrated in space in Figure~\ref{Fig_surf2de} and in space-time in Figure~\ref{Fig_hyprot}, taken from simulations using the Particle-In-Cell code Warp \cite{Warp}. 
The Lorentz transformation can be described as a hyperbolic rotation in Minkowski space and its rotational effect is explicitly visible in Figure~\ref{Fig_hyprot}.

Figure~\ref{Fig_surf2de} shows surface renderings of the transverse and longitudinal electric fields respectively, as the beam enters its early stage of acceleration by the plasma wake, from calculations in the laboratory frame and a Lorentz boosted frame at $\gamma=13$ (approximately the laser group velocity in the plasma column $\gamma_g\approx13.2$). The two snapshots offer strikingly different views of the same physical processes: in the laboratory frame, the wake is fully formed before the beam undergoes any significant acceleration, the laser (Fig. \ref{Fig_surf2de} top-left) is easily recognizable (i.e. its shape is only slightly distorted by the plasma) and leaves a visible imprint on the wake (longitudinal) field (Fig. \ref{Fig_surf2de} bottom-left); in the boosted frame, the beam is accelerated as the plasma wake develops, the laser (Fig. \ref{Fig_surf2de} top-right) is not easily recognizable (i.e. its shape is highly distorted by the plasma) and no evident imprint is left on the wake field (Fig. \ref{Fig_surf2de} bottom-right).

The physics underlying the differences between $\gamma=1$ and $\gamma=13$ views of the wake is illustrated by histories of the (transverse) laser field on the longitudinal axis, reported in Figure~\ref{Fig_hyprot} from simulations performed using the laboratory frame and boosted frames at $\gamma=5$ and $13$. Simulations with $\gamma=1$ and 5 used a moving window propagating at the group velocity $v_g$ of the laser in the plasma and the data are plotted in the frame of the (galilean) moving window. The simulation with $\gamma=13$ did not use a moving window as it was performed very near the group velocity of the laser in the plasma $\gamma_g\approx13.2$.  The data from the boosted frame simulation at $\gamma=13$ is presented in the boosted frame as well as in the laboratory frame moving window (after Lorentz transformation), allowing direct comparison with the simulation calculating with the laboratory frame. The calculation with the boosted frame at $\gamma=13$ was approximately 200 times faster than the calculation with the laboratory frame, as expected \cite{VayPRL07}. The agreement between the two is nonetheless excellent (comparing top-left plot to bottom-right plot in Figure~\ref{Fig_hyprot}), confirming the accuracy of the calculation in the boosted frame. The group velocity of the wake in the plasma is always below the speed of light in vacuum while the phase velocity is always above it, resulting in oblique stripes in the laboratory frame plot. As $\gamma$ boost rises, the stripes rotate according to the rules of the Lorentz transformation, eventually becoming nearly perpendicular to the time axis as $\gamma$ boost nears $\gamma_g$ (bottom-left plot in Figure~\ref{Fig_hyprot}), the laser group velocity approaches zero and the phase velocity approaches infinity. In effect, the laser oscillations that appear in the laboratory as spatial oscillations propagating in the plasma are transformed into time beating of the field for calculations in frames whose boost nears the laser group velocity. As discussed below, this effect has important consequences for the modeling of full scale stages at 10 GeV or above.

\textit{Mitigation of a numerical instability.---}
Several numerical limits have restricted the boost performance in past simulations: laser initialization, statistics and a short wavelength instability.
The first two are discussed elsewhere \cite{VayAAC10} and we concentrate here on the latter.

A violent high-frequency numerical instability developing at the front of the plasma column for boosts at $\gamma \gtrsim 100$ in 2D and $\gamma \gtrsim 50$ in 3D was reported by various authors  \cite{VayDPF09,MartinsCPC10,BruhwilerPC08}. The presence and growth rate of the instability was observed to be very sensitive to the resolution (slower growth rate at higher resolution), to the amount of damping of high frequencies, to smoothing of short wavelengths, and to the boost value (stronger instability at higher boost). An extensive set of testing was performed with Warp to investigate simulations of downscaled 100  MeV and full scale 10 GeV LPA stages \cite{VayARXIV10}. One of the key findings was that frames with higher boosts (up to $\gamma_g$) allow for higher levels of filtering and damping than is possible in other frames for the same accuracy, allowing mitigation of the instability. This is a direct  consequence, and benefit, of the hyperbolic rotation effect from the Lorentz boost seen in Figure~\ref{Fig_hyprot}.

\begin{figure}[htb]
\centering
{\small
\begin{tabular}{@{}c@{}c@{}}
  Lab ($\gamma=1$) &
  Boost ($\gamma=13$) \\
  \includegraphics*[height=0.19\textwidth]{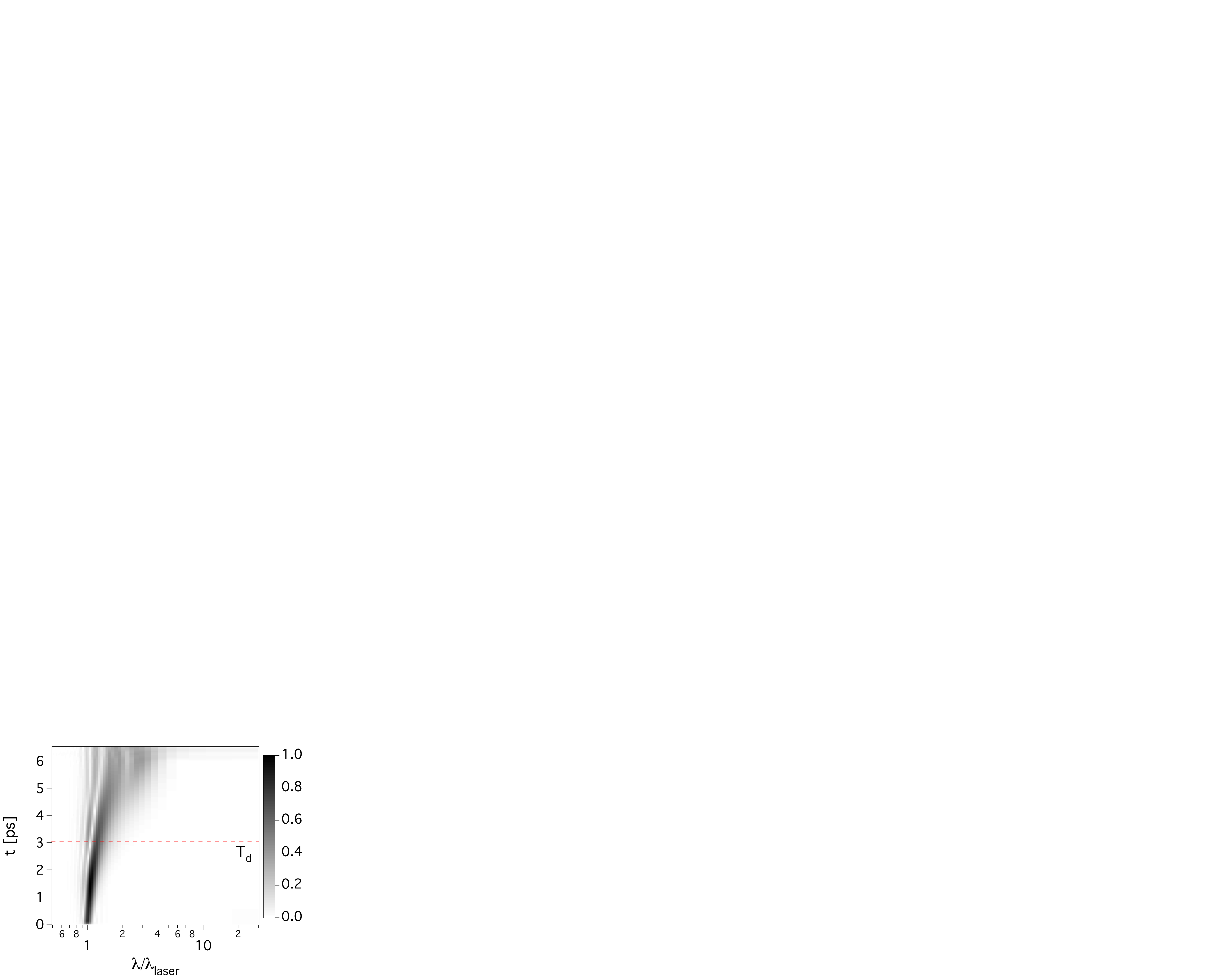} &
  \includegraphics*[height=0.19\textwidth]{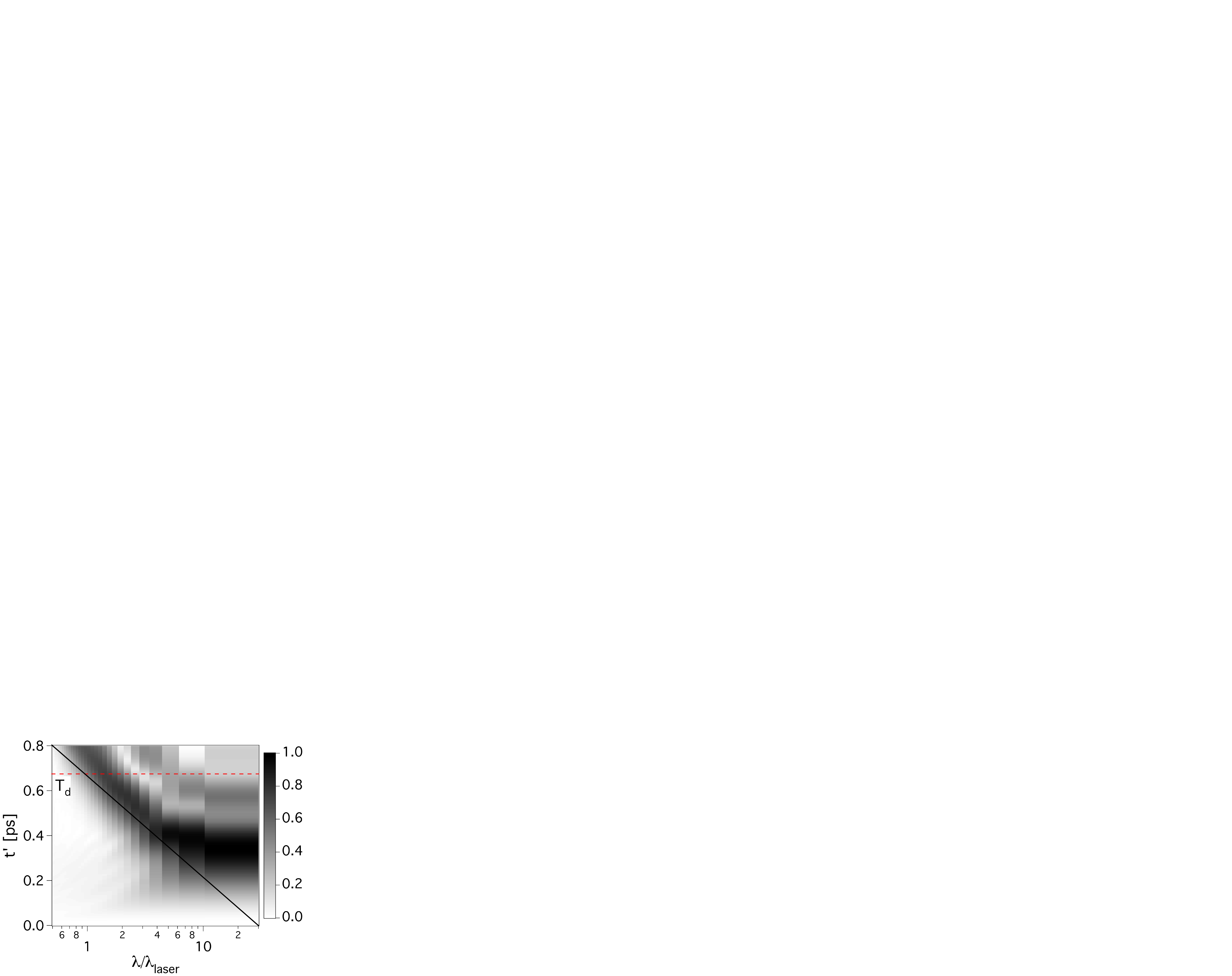} \\
  Lab ($\gamma=1$) &
  Boost ($\gamma=130$)\\
  \includegraphics*[height=0.19\textwidth]{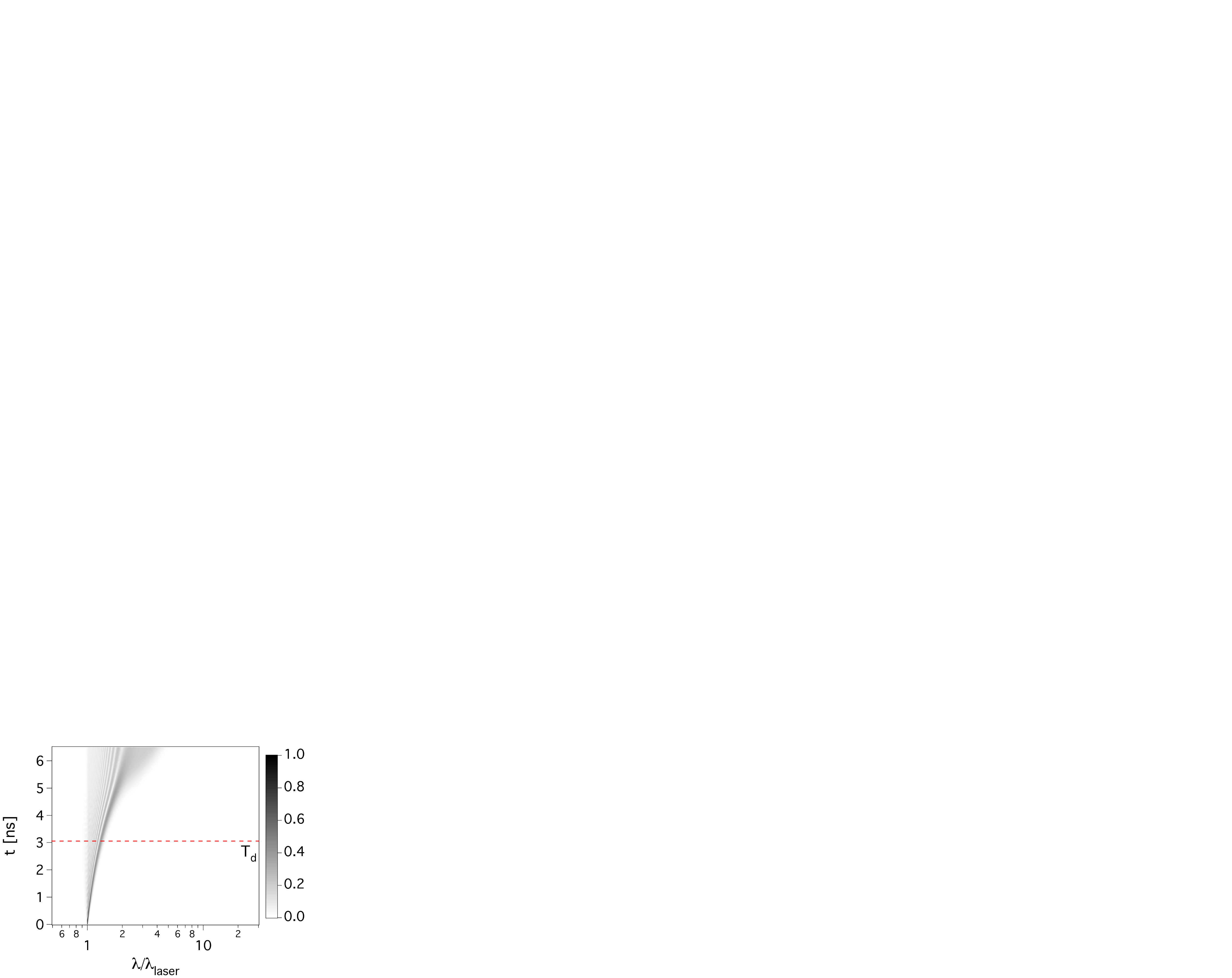} &
  \includegraphics*[height=0.19\textwidth]{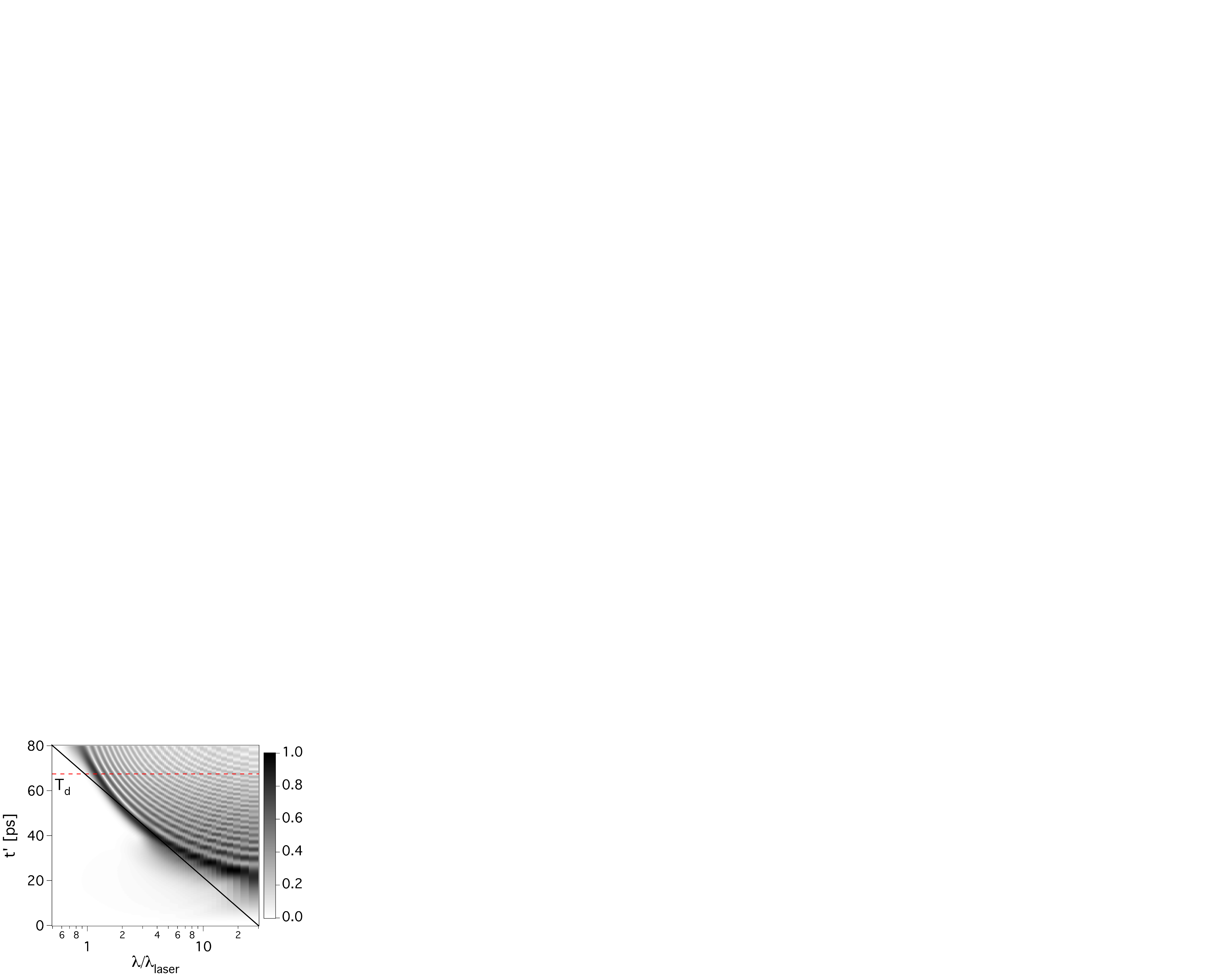} 
\end{tabular}
}
  \caption{(color online) Spectral content history of the laser field on axis as it propagates through the plasma column for 100 MeV (top) and 10 GeV (bottom) stages, in the laboratory frame (left) and near the laser group velocity frames at $\gamma=13$ (top-right) and $130$ (bottom-right).}
\label{Fig_spectrum}
\end{figure}

The spectral content history of the laser field on axis as it propagates through the plasma column is given in Figure~\ref{Fig_spectrum} for 100 MeV and 10 GeV stages, in the laboratory frame and in the laser group velocity frames at $\gamma=13$ and $130$ respectively. Laser depletion occurs at times $T_d=L_d/c$ at respectively $T_d\approx3.1$ ns and  $T_d\approx3.1$ ps for 100 MeV and 10 GeV stages.
In the laboratory frame, the spectral content of the laser is concentrated initially in a narrow band around the nominal laser wavelength, then spreads due to dispersion and depletion effects as the laser propagates through the plasma. In the frame of the laser group velocity, much of the spectral content is localized initially at wavelengths that are several times the nominal laser wavelength (in vacuum), then progressively fills lower portions of the spectrum. 
As in practice the longitudinal numerical resolution is set relative to the vacuum laser nominal wavelength, a higher level of filtering (or damping) is acceptable for simulations in a boosted frame than for those in the laboratory frame. 
Furthermore, the higher the boost, the longer the wavelengths with substantial spectral content in the early part of the laser propagation (comparing spectral content below diagonal in plots from right column of Figure~\ref{Fig_spectrum}), meaning that higher levels of filtering are allowable at higher boosts were the growth of the numerical instability is the strongest. 

\begin{figure}[htb]
   \centering
    \includegraphics*[width=70mm, trim=0.in 1.8in 0.in 0.in]{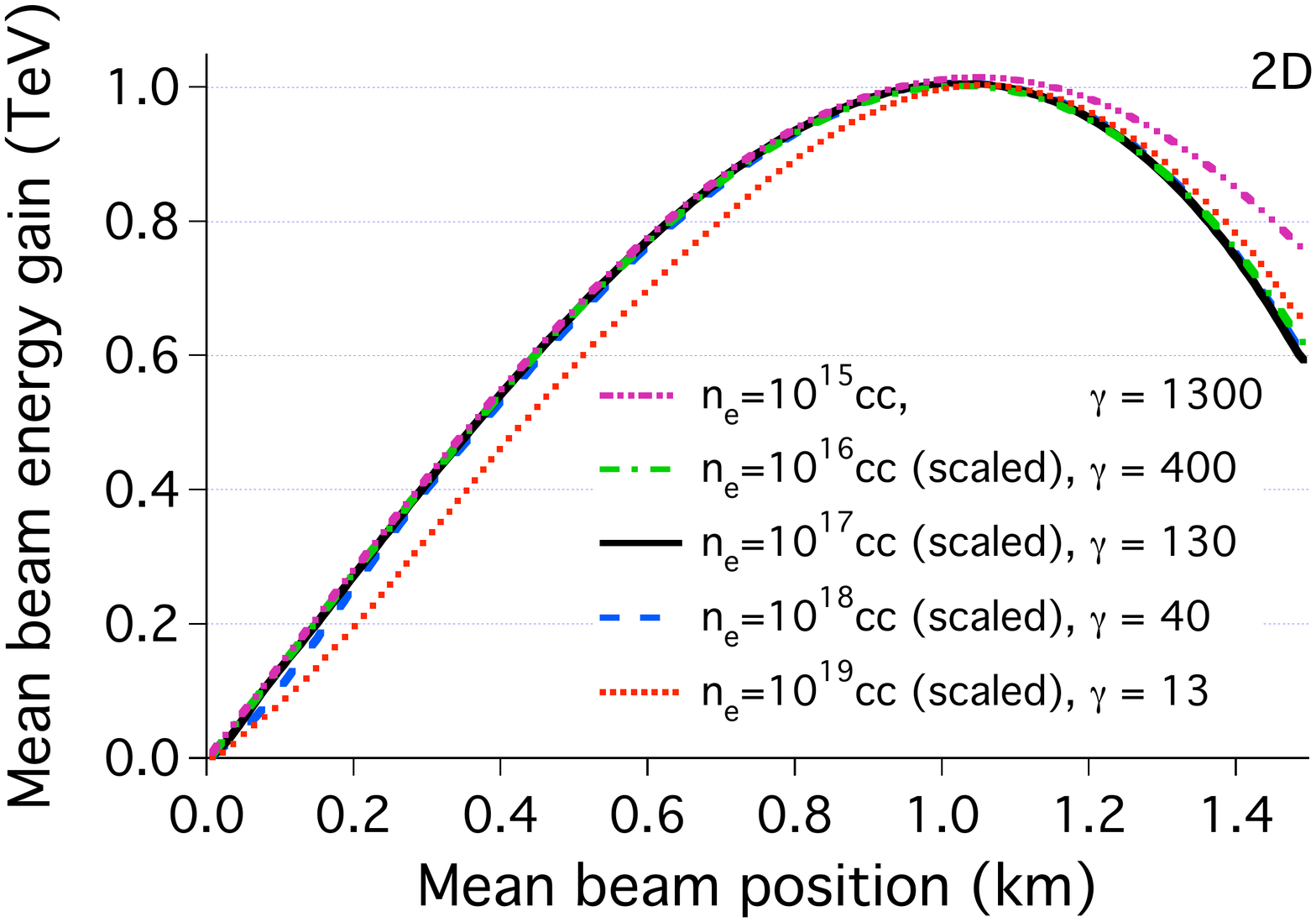} 
    \includegraphics*[width=70mm, trim=0.in 1.8in 0.in .in]{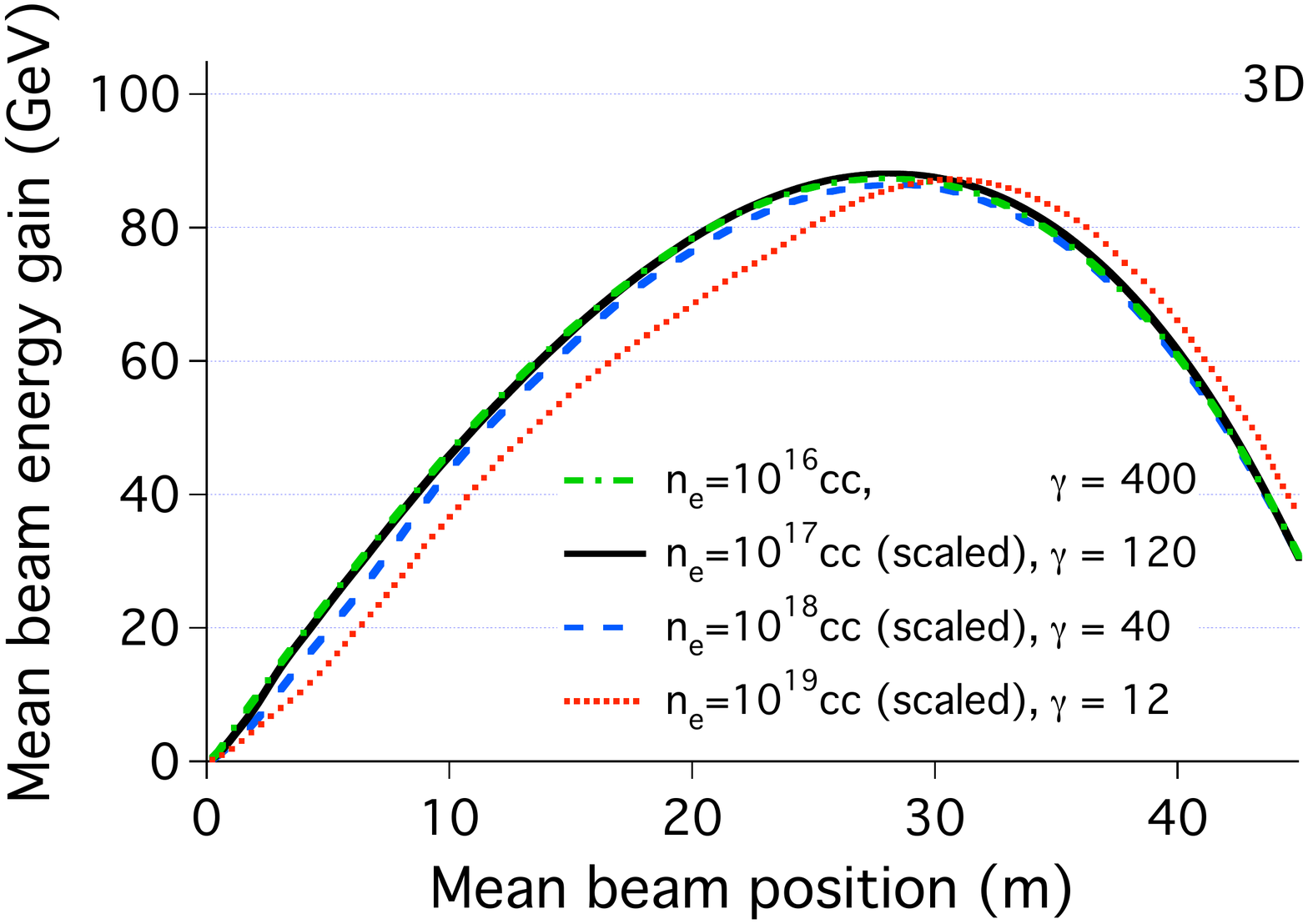} 
   \caption{(color online) Verification of the scaling \cite{CormierAAC08,GeddesPAC09} of electron beam energy gain versus longitudinal position (in the laboratory frame) from direct simulations at $n_e=10^{19}$ cc down to $10^{15}$ cc (energy gains from 0.1 GeV to 1 TeV), using Lorentz boosted frames of reference at $\gamma$ between $12$ and $1300$, in 2-1/2D (top) and 3D (bottom). Energies from the lowest energy stages were scaled to the highest energy stage to allow for direct comparison.}
   \label{Fig_ehist100GeV}
\end{figure}
\textit{Modeling of  up to 1 TeV stages.---}
Mitigation of the instability allowed simulations of stages with gain energy as high as 1 TeV in 2-1/2D and 
100 GeV in 3-D, using Lorentz boosted frame with $\gamma$ as high as $1,300$ (see Figure~\ref{Fig_ehist100GeV}), offering for the first time direct verification of the scaling of plasma accelerators into the 1 TeV range for deeply depleted stages.
Simulations with the highest boost necessitated filtering of short wavelength over a wider band, and in agreement with the observations of the previous section, the accuracy was not compromised. 
The highest level of smoothing was needed for the 1~TeV case, explaining the deviation past 1 km. This deviation is of little importance in
practice, where one is mostly interested in the beam evolution up-to the peak energy point. The differences at $n_e=10^{19}$cc can be attributed to the effects from having only a few laser oscillations per pulse. 
The theoretical speedup \cite{VayAAC10} of the full scale 100 GeV class run is estimated to be over 100,000. Assuming the use of a few thousands of CPUs, a simulation that would have required an impractical several decades to complete using the laboratory frame, was completed in only four hours using 2016 CPUs of the Cray system at NERSC. The speedup of the 2-1/2D 1~TeV stage is estimated to be over a million. 

\begin{figure}[htb]
   \centering
    \includegraphics*[width=85mm, trim=0.in 1.8in 0.in 0.in]{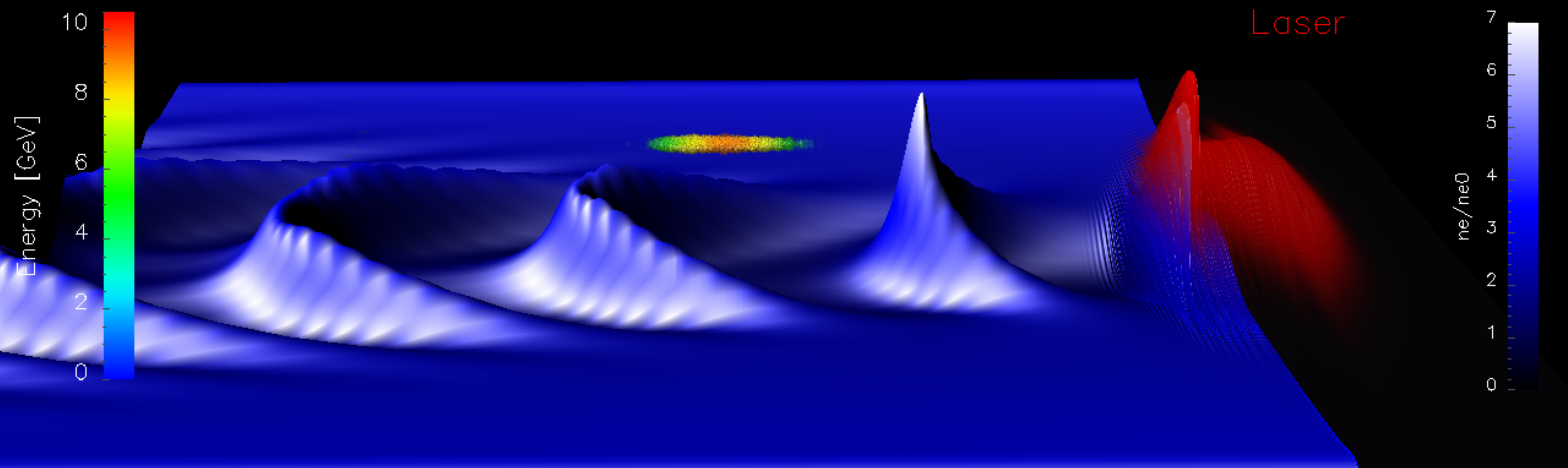} 
   \caption{(color online) Snapshot from a 10 GeV LPA stage boosted frame simulation. The image shows an externally injected electron bunch (rainbow color) riding a density wake (blue) excited by an intense laser pulse (red), propagating in a 0.65 m long plasma channel. The laser pulse (~40 J in ~67 fs), focused to ~90 $\mu$m spot size at the entrance
of the channel, has reached the end of the plasma channel. The electron bunch energy has reached up to $\sim$10 GeV.}
   \label{Fig_bella}
\end{figure}

The boosted frame Particle-In-Cell  technique accurately resolves  the wavelength shifting and
broadening that occurs as the laser depletes, offering advantages over other  models (for example envelope, quasistatic)  while providing the speed required for direct simulation of 10 GeV and beyond laser plasma accelerators to accurately model laser and beam transverse oscillations. It is being
applied to the direct simulation of 10 GeV beam loaded
stages for detailed designs of experiments on new lasers
such as BELLA \cite{LeemansAAC10} (see Figure~\ref{Fig_bella}), as well as next generation
controlled laser plasma accelerator stages and collider modules \cite{SchroederAAC08}.

In summary, direct simulations of stages in the range of 0.1 GeV-
1 TeV have been performed using the Lorentz boosted
frame technique, verifying for the first time the performance of plasma accelerators into the 1 TeV range for
deeply depleted stages. This has been possible thanks to
effects of the hyperbolic rotation in Minkowski space of
the laser propagation in the plasma column which has
been key, in conjunction with the development of novel
numerical techniques (described elsewhere), in allowing
successful mitigation of a violent numerical instability
that was limiting the boost performance in past simulations.
As a result, the maximum theoretical
speedup of over a million for a 1 TeV stage was realized, which is three
orders of magnitude higher than was possible previously.
The new developments offer unique highly efficent tools
for the detailed designs of experiments on new lasers such
as BELLA \cite{LeemansAAC10}.


\begin{acknowledgments}
Work supported by US-DOE Contracts DE-AC02-05CH11231 and US-DOE SciDAC program ComPASS. Used resources of NERSC and LBNL cluster Lawrencium, supported by US-DOE Contract DE-AC02-05CH11231. The authors thank D.~L. Bruhwiler, J.~R. Cary, E. Esarey, A. Friedman, W.~P. Leemans, S.~F. Martins, W.~B. Mori, R.~D. Ryne and C.~B. Schroeder for insightful discussions and/or advice in preparing this manuscript.
 \end{acknowledgments}



\bibliography{PRL_vay_2010}

\IfFileExists{\jobname.bbl}{}
 {\typeout{}
  \typeout{******************************************}
  \typeout{** Please run "bibtex \jobname" to optain}
  \typeout{** the bibliography and then re-run LaTeX}
  \typeout{** twice to fix the references!}
  \typeout{******************************************}
  \typeout{}
 }

\end{document}